\documentclass{sf2a-conf2014}
\usepackage{amsmath,amsthm, amsfonts,multirow}
\usepackage{graphicx}
\usepackage{hyperref}
\usepackage[]{natbib}  
\usepackage{epstopdf}

\def\BibTeX{{\rm B\kern-.05em{\sc i\kern-.025em b}\kern-.08em
    T\kern-.1667em\lower.7ex\hbox{E}\kern-.125emX}}
\bibpunct{(}{)}{;}{a}{}{,}  

\usepackage{color}
\definecolor{darkgreen}{RGB}{0,142,128}

\newcommand{\lrsp}{\emph{Liv. Rev. Sol. Phys.}}

\begin{document}

\TitreGlobal{SF2A 2014}


\title{Modelling the Corona of HD 189733 in 3D}

\runningtitle{Modelling the Corona of HD 189733 in 3D}

\author{A. Strugarek}\address{D\'epartement de physique, Universit\'e
  de Montr\'eal, C.P. 6128 Succ. Centre-Ville, Montr\'eal, QC H3C-3J7,
  Canada}

\author{A. S. Brun}\address{Laboratoire AIM Paris-Saclay, CEA/Irfu Universit\'e Paris-Diderot CNRS/INSU, F-91191 Gif-sur-Yvette.}

\author{S. P. Matt}\address{Astrophysics group, School of Physics, University of Exeter, Stocker Road, Exeter EX4 4QL, UK }

\author{V. Reville$^{2}$}

\author{J. F. Donati}\address{Laboratoire d'Astrophysique, Observatoire Midi-Pyr\'en\'ees, 14 Av. E. Belin, F-31400 Toulouse, France}

\author{C. Moutou}\address{Aix-Marseille Universit\'é, CNRS, LAM (Laboratoire d’Astrophysique de Marseille) UMR 7326, 13388 Marseille, France}

\author{R. Fares}\address{SUPA, School of Physics and Astronomy, University of St Andrews, North Haugh, St Andrews KY16 9SS, UK}

\setcounter{page}{237}


\maketitle


\begin{abstract}
The braking of main sequence stars originates mainly from their
stellar wind. The efficiency of this angular momentum extraction
depends on the rotation rate of the star, the acceleration profile of
the wind and the coronal magnetic field. The derivation of scaling
laws parametrizing the stellar wind torque is important for
our understanding of gyro-chronology and the evolution of the rotation
rates of stars. In order to understand the impact of complex magnetic
topologies on the stellar wind torque, we present 
three-dimensional, dynamical simulations of the corona of HD
189733. Using the observed complex topology of the magnetic field, we
estimate how the torque associated with the wind scales with model
parameters and compare those trends to previously published scaling
laws.  
\end{abstract}



\section{Introduction}

Magnetized stellar winds have long been recognized as the major source of
angular momentum extraction in main sequence stars
\citep{Parker:1958dn,Weber:1967kx,Mestel:1968aa}. In order to reliably
assess the stellar wind torque, the acceleration
profile and the magnetic field geometry of the wind are required. It was recently
demonstrated that, in particular, complex magnetic topologies
of cool stars have a major impact on the torque \citep[see,
\textit{e.g.}][]{Cohen:2014et,Reville:2014ud} compared to more
simple topologies. Three dimensional
numerical simulations provide a reliable way to compute, in a
dynamically self-consistent way, the torque arising from stellar wind
with complex magnetic fields. However, no parametrization
of fully three-dimensional,
non-axisymmetric stellar wind torques has yet been proposed in the
literature.


We report here an ongoing effort in developing
magnetohydrodynamics (MHD) simulations of the stellar winds of cool
stars in three dimensions using complex magnetic field topologies. We
consider one-fluid and
ideal models of stellar winds which are simple compared to
the most recent solar wind models \citep[see,
\textit{e.g.},][]{Oran:2013aa,Sokolov:2013ki}. However, they inherit
important conservation properties from their 2.5D counterparts
\citep[see][]{Strugarek:2012th,Strugarek:2014tf} which makes them
reliable to derive general scaling laws for the stellar wind breaking.

We focus, in this proceeding, on the extension of our three-dimensional
stellar wind model \citep[see][]{Strugarek:2014tf}
to take into account arbitrarily complex magnetic topology. As a
test-bench we use the
case of HD 189733 (see Section \ref{sec:spectr-maps-from}), which has
already been modelled in 3D by \citet{Llama:2013il}. We vary one of the free
parameters of the model, the Alfv\'en speed at the base of the corona
(Section \ref{sec:stellr-wind-modell}), and compare our results with
previously published results. Furthermore, the various cases presented
here allow us to compare the results of fully three-dimensional
simulations with scaling laws that
were derived in a magnetic topology-independent manner in axisymmetric geometry
\citep{Reville:2014ud}. We find a good general agreement with this
law, though the predicted torque is found to be generally larger than
the one we found with our 3D models. Given that the scaling law was derived in
axisymmetric geometry (in 2D), we find the general agreement
satisfactory and are encouraging for further exploration of this scaling law, using
fully three-dimensional numerical simulations. 


\section{Characteristics of HD 189733}
\label{sec:spectr-maps-from}

The properties of the planet-hosting HD 189733 star have been reported
in \citet{Bouchy:2005aa}. It is a K2V star with a mass $M_{\star} =
0.82 \pm 0.03 M_{\odot}$ and a radius $R_{\star} = 0.76 \pm 0.01
R_{\odot}$ \citep{Winn:2007aa}. The rotation period of HD 189733 has
been characterized by several teams using photometry
\citep{Hebrard:2006aa,Winn:2007aa}, and is reported between $11.8$ and $13.4$
days. Here we adopt a rotation period of $12$ days.

\begin{figure}[htbp]
  \centering
  \includegraphics[width=0.5\linewidth]{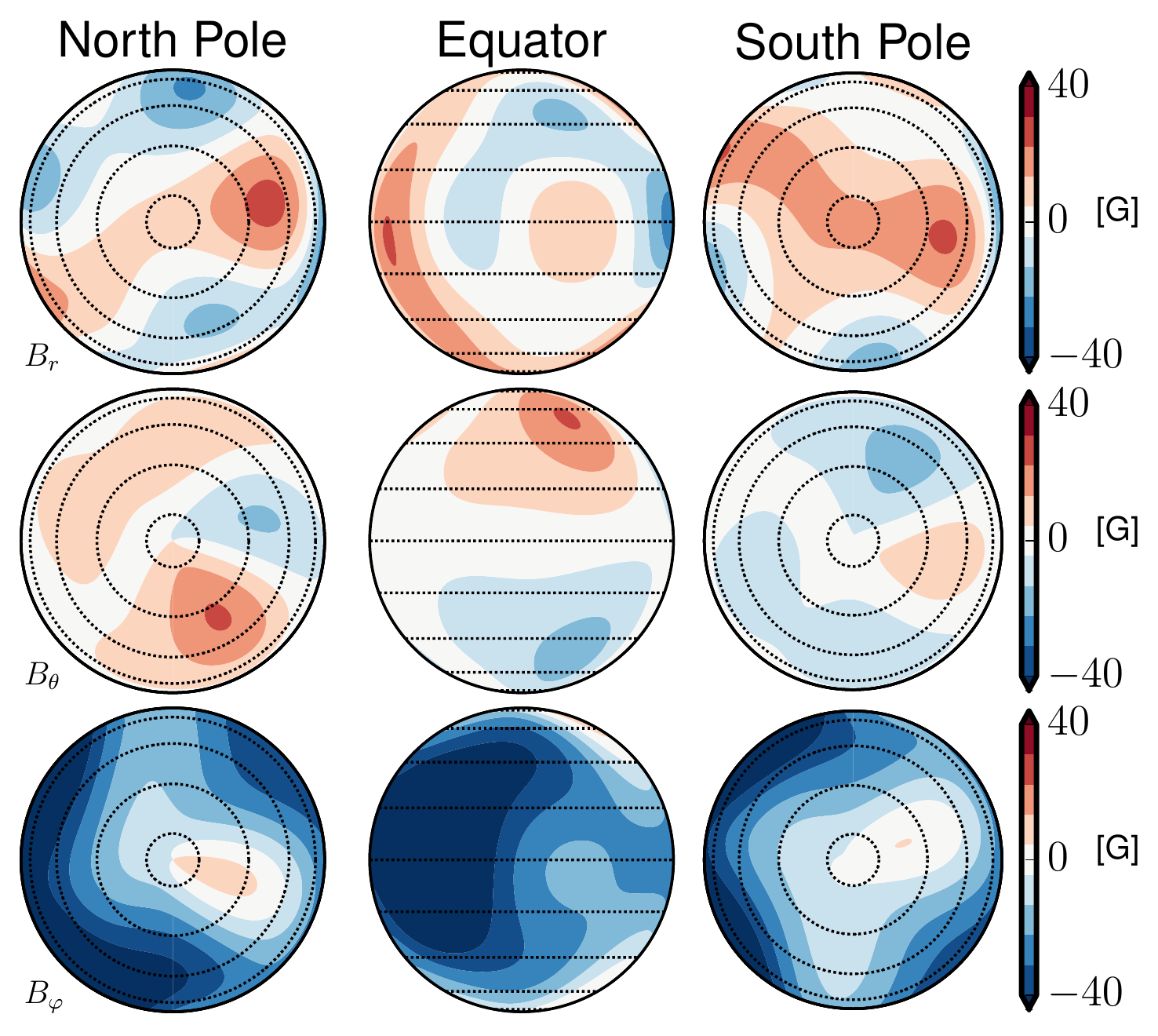}
  \caption{Spectro-polarimetric reconstruction of the coronal magnetic field of
    HD 189733 in July 2008 \citep[see][]{Fares:2010hq}. The amplitude
    of the field is given in Gauss.\label{fig:maps}}
\end{figure}

We use the spectro-polarimetric magnetic maps of HD 189733 obtained by
\citet{Fares:2010hq} that where observed in July 2008 with NARVAL. We
show in figure \ref{fig:maps} the reconstructed components of the
magnetic field in spherical coordinates (top to bottom) in
orthographic projection viewed from the north pole (left), the equator
(middle) and the south pole (right). The magnetic field is highly
non-axisymmetric and derives from 20 spherical harmonics modes ($l_{max}=5$).

\section{Modelling the wind of HD 198733}
\label{sec:stellr-wind-modell}

\subsection{Numercial Model and Parameters}
\label{sec:numercial-model}

Following the work in 2.5D axisymmetric geometry described by
\citet{Strugarek:2014ab} and in 3D by \citet{Strugarek:2014tf}, we adapted our stellar wind
model to account for the magnetic topology of observed stars. 
We use the PLUTO code \citep{Mignone:2007iw} which solves the following set of ideal MHD equations:
\begin{eqnarray}
  \label{eq:mass_consrv_pluto}
  \partial_t \rho + \boldsymbol{\nabla}\cdot(\rho \mathbf{v}) &=& 0 \, \\
  \label{eq:mom_consrv_pluto}
  \partial_t\mathbf{v} +
  \mathbf{v}\cdot\boldsymbol{\nabla}\mathbf{v}+\frac{1}{\rho}\boldsymbol{\nabla} P
  +\frac{1}{\rho}\mathbf{B}\times\boldsymbol{\nabla}\times\mathbf{B}
  &=& \mathbf{a} \, ,
  \\
  \label{eq:ener_consrv_pluto}
  \partial_t P +\mathbf{v}\cdot\boldsymbol{\nabla} P + \rho
  c_s^2\boldsymbol{\nabla}\cdot\mathbf{v} &=& 0 \, ,\\
  \label{eq:induction_pluto}
  \partial_t \mathbf{B} - \boldsymbol{\nabla}\times\left(\mathbf{v}\times\mathbf{B}\right)
  &=& 0 \, ,
\end{eqnarray}
where $\rho$ is the plasma density, $\mathbf{v}$ its velocity, $P$ the gas
pressure, $\mathbf{B}$ the magnetic field, and $\mathbf{a}$ 
is composed of the gravitational acceleration (which is
time-independent) and of the Coriolis and centrifugal forces. The
equations are solved in a frame rotating at the stellar rotation rate $\Omega_{\star}$.
The sound speed is given by $c_s=\sqrt{\gamma\,P/\rho}$, with
$\gamma$ the ratio of specific heats. We use an
ideal gas equation of state
\begin{equation}
  \label{eq:EOS}
  \rho\varepsilon = P/\left(\gamma-1\right)\, ,
\end{equation}
where $\varepsilon$ is the specific internal energy per mass. We use an \textit{hll}
solver combined with a \textit{minmod} limiter. A
second-order Runge-Kutta is used for the time evolution, resulting in
an overall second-order accurate numerical method. The solenoidality
of the magnetic field is ensured with a constrained transport
method \citep[see][]{Mignone:2012aa}. We refer the interested reader to
\citep{Mignone:2007iw} for an extensive description of the various
numerical methods that PLUTO offers. For this first study, we use a
low-resolution grid of 224$^{3}$ points for a domain size of
$[-20,20]^3$. 96$^{3}$ uniform grids points are use to discretize the
$[-1.5,1.5]^{3}$ domain and stretched grids are used elsewhere.

The first control parameter of the modelled stellar wind is the ratio
of specific heats $\gamma$. In order to ensure that the MHD modelling
the wind produces velocities compatible with solar inferences (of the
order of $400$-$500$ km s$^{-1}$ at $1$~AU in the case the Sun), we
choose a close-to-isothermal value of $\gamma=1.05$. 
The structure of the stellar wind is then controlled by a set of
parameters that can be written as velocity ratios at the surface of
the star \citep[see, \textit{e.g.},][]{Matt:2012ib}.

In this work we model the corona of HD 189733. The escape velocity at
the surface of HD 189733 is $v_{\rm esc} = 6.41 \times 10^{7}$ cm/s. 
We choose a standard coronal temperature of $2 \times 10^6$ K
\citep[equivalent to][]{Llama:2013il} that
sets the normalized sound speed at the surface of the star to
$c_{s}/v_{\rm esc} = 0.29$. The rotation period of the star is $12$
days, which results in a normalized rotation velocity $v_{\rm rot}/v_{\rm esc} =
5.0 \times 10^{-3}$. Based on these parameters, the stellar wind is then driven by our 
boundary conditions representing the base of the corona \citep[see][for
complete discussions on those boundary conditions]{Strugarek:2014fr,Strugarek:2014ab}. 

The last parameter controlling the stellar wind is the
magnetic field. We use the radial component of the observed magnetic
field (Figure \ref{fig:maps}) and perform a potential extrapolation
to define the 3D magnetic field in the whole domain at initialization
\citep[see the Appendix C in][for a full description of the potential
extrapolation technique]{Schrijver:2003vu}. The magnetic field in the
boundary condition --which is effectively impacting the wind driving--
depends only marginally on the exact location of the source
surface used for the potential extrapolation. The initial magnetic
field in the whole domain is directly
related to the location of the source surface, but this initial
condition is rapidly modified by the wind that establishes a
steady-state configuration with both open and closed field regions
that is independent of the initial potential field. As a result, the
choice of the source
surface has no impact on our simulation results as long as it is sufficiently
distant from the stellar surface.

The amplitude of the magnetic field is constrained by the
observations (Figure \ref{fig:maps}), but the density at the base of
the corona is not. Hence, the Afl\'ven speed at the base of our model
is not well constrained. It can generally
be related to the mass loss rate induced by the
stellar wind \citep[see][]{Matt:2012ib,Reville:2014ud}, which in some cases can be deduced from observations of
astrospheric Ly$\alpha$ absorption \citep{Wood:2004uc}. However, we lack
such observations for HD 189733 and are thus compelled to test a range
of Alfv\'en speeds. We hereafter define the
maximum Alfv\'en speed at the base of the corona by $v_{a} = {\rm max} (B) /
\sqrt{4\pi\rho_{\star}}$, where $B$ is the total magnetic field
amplitude and $\rho_{\star}$ the density at the base of the
corona. $B$ is constrained by the observations,
hence we choose three different base density values $\rho_{\star}\in
\left\{1.51\times
  10^{-14},\, 2.13\times 10^{-15},\, 8.3\times 10^{-16} \right\}$ g
cm$^{-3}$ that lead to the averaged velocity ratios $v_{a}/v_{\rm esc}
\in \left\{2.58,\, 6.79,\, 11.0 \right\}$. The highest density
  we chose corresponds to the base density value chosen by
  \citet{Llama:2013il} which was tuned to reproduce the observed X-ray
  luminosity of HD 189733.

\subsection{Results}
\label{sec:results}

\begin{figure}[htbp]
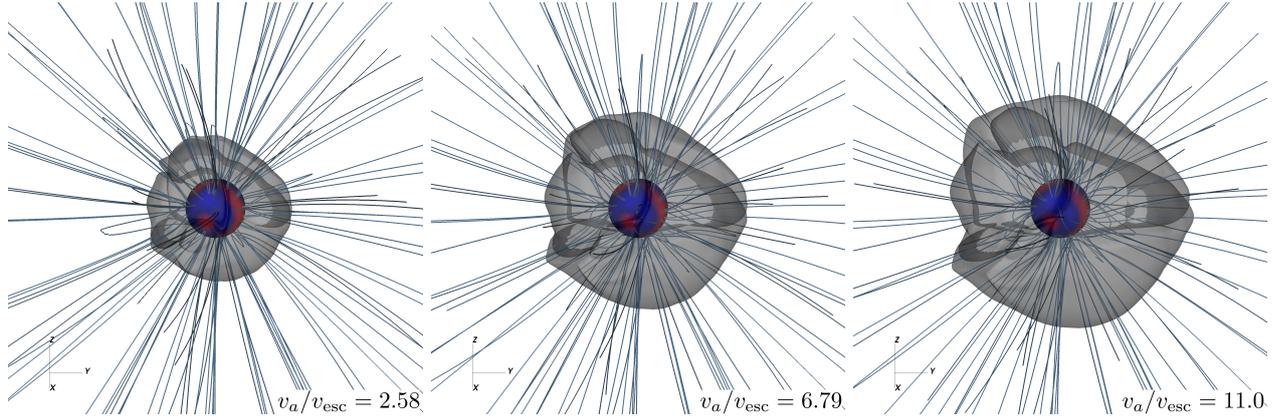

  \centering
  \includegraphics[width=0.32\linewidth]{strugarek_figure2}
  \includegraphics[width=0.32\linewidth]{strugarek_figure3}
  \includegraphics[width=0.32\linewidth]{strugarek_figure4}
  \caption{3D renderings of the coronal magnetic field (blue lines)
    of HD 189733. The color sphere shows the radial component of the
    magnetic field at the surface of the star. The transparent blue
    surface labels the Alfv\'en surface of the wind. From left to
    right, $v_{a}/v_{\rm esc}$ is $2.58$, $6.79$, and $11.0$.\label{fig:3D_views}}
\end{figure}

In Figure \ref{fig:3D_views} we show three dimensional renderings of
the dynamical corona of HD 189733 for the three Alfv\'en velocities we
considered. The blue (negative) and orange (positive) map shows the radial magnetic field at
the base of the corona, where the stellar wind is driven in our
model. The coronal magnetic field lines are shown in light blue, and
the Alfv\'en surface is labelled by the transparent grey surface. We
observe that for increasing base Alv\'en velocity, the Alfv\'en
surface is further away from the star, retaining a similar global
shape. This is a simple consequence of the decrease of the base
density in the corona: the magnetic tension strengthens compared to
the intertial and pressure forces of the wind, and the magnetic field retains more closed
field lines, where the coronal plasma is imprisoned and the stellar wind
hindered. 

We can estimate \textit{a posteriori} the mass loss rate $\dot{M}_{w}$ and
angular momentum loss rate $\dot{J}_{w}$ of the stellar
winds. In the third case ($v_{a}/v_{\rm esc}=11$), we find a mass loss
rate of $4.35 \times 10^{-13} \, M_{\odot}$ yr$^{-1}$, which is very close
to the mass loss rate of $4.5 \times 10^{-13}
\, M_{\odot}$ yr$^{-1}$ found by \citet{Llama:2013il}. The two other cases have higher mass loss
rates, respectively $8.66\times 10^{-12}$  and $1.16\times 10^{-12}$
$M_{\odot}$ yr$^{-1}$. Following
\citet{Matt:2012ib,Reville:2014ud}, we
define a torque-derived, effective Alfv\'en radius by relating the two loss rates
through
\begin{equation*}
  \frac{R_{a}}{R_{\star}} \equiv
  \sqrt{\frac{-\dot{J}_{w}}{\dot{M}_{w}\Omega_{\star}}} \, ,
\end{equation*}
where $\Omega_{\star}$ is the rotation rate of the star. It was
recently shown by \citet{Reville:2014ud} that this effective
Alfv\'en radius --at least in 2.5D axisymmetric models-- can be
related to the wind magnetization parameter
\begin{equation*}
  \Upsilon_{o} \equiv
  \frac{\Phi_{o}^{2}}{R_{\star}^{2}\dot{M}_{w}v_{\rm esc}} \, ,
\end{equation*}
where the open magnetic flux is given by
\begin{equation*}
  \Phi_{o} = \int_{S} \left|\mathbf{B}\cdot{\rm d}\mathbf{S} \right|
  \, ,
\end{equation*}
where $\int_{s} . {\rm d}\mathbf{S}$ stand for the integral over a
closed surface at a sufficiently large distance from the central star
for the magnetic field lines to be all open. The average Alfv\'en
radius and the magnetic confinement parameter are
related through
\begin{equation}
  \label{eq:scaling_law}
  \frac{R_{a}}{R_{\star}} = K_{3} \left[ \frac{\Upsilon_{o}}{\sqrt{1 +
      (f/K_{4})^{2}}}\right]^{m}\,
\end{equation}
where the coefficients $K_{3}$, $K_{4}$ and $m$ were determined
empirically by \citet{Reville:2014ud} to be $K_{3} = 1.4\pm 0.1$,
$K_{4} = 0.06 \pm 0.01$ and $m=0.31\pm 0.02$. 

\begin{figure}[htbp]
  \centering
  \includegraphics[width=0.6\linewidth]{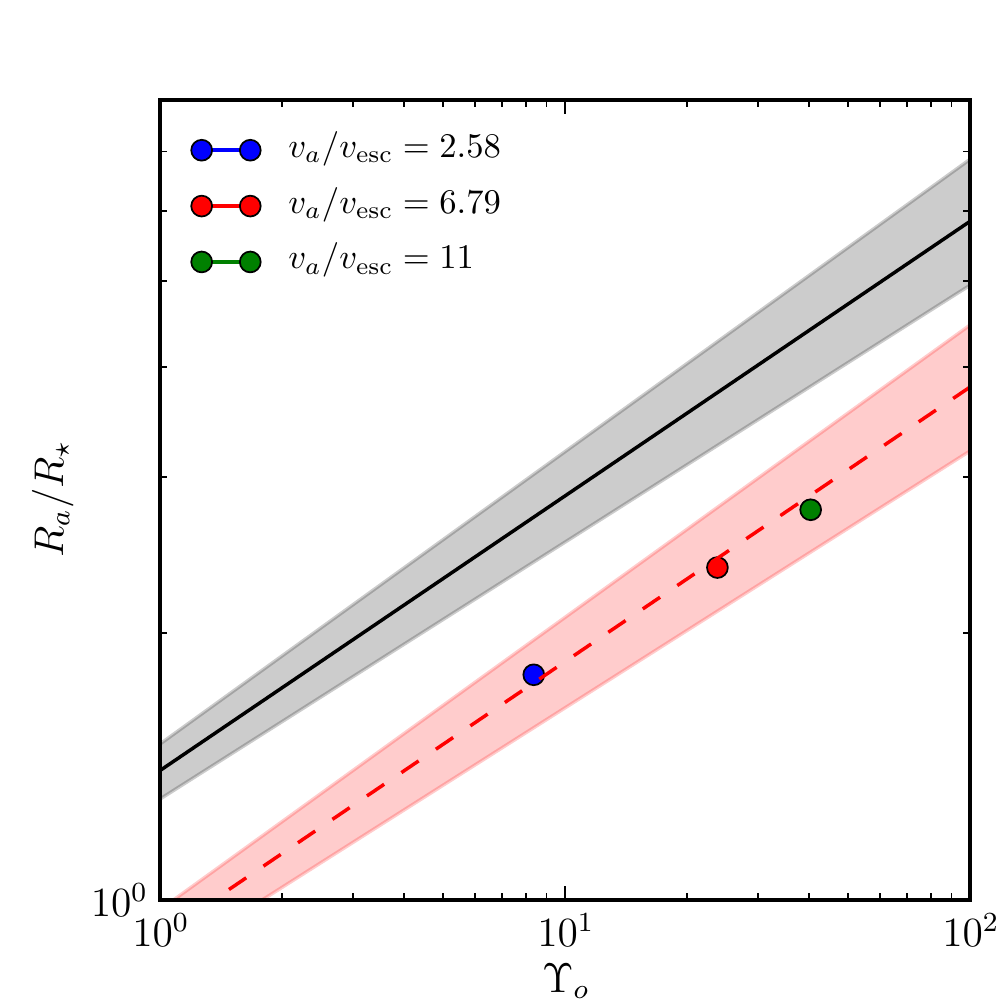}
  \caption{Generalized Alfv\'en radius as a function of the wind magnetization
    parameter $\Upsilon_{o}$ (see text). The black line
    and grey area show the scaling law prediction from
    \citet{Reville:2014ud}. The dashed red line and red area represent the
    same scaling law, modified to account a different sound
    speed at the base of the corona.\label{fig:compare_breaking_law}}
\end{figure}

We compute $R_{a}$ and $\Upsilon_{o}$ in our three models and display
them in Figure \ref{fig:compare_breaking_law}. The scaling law
predicted by \citet{Reville:2014ud} is shown by the black line. The
grey area labels the error bars of this scaling law. Our
cases seem to follow roughly the scaling-law
trend, but appear to be shifted downwards. This is in reality awaited
since our reference scaling law was derived with a fixed sound speed
ratio of $c_{s}/v_{esc}=0.222$ \citep[see][]{Reville:2014ud}. In our
case, we consider a much larger sound speed $c_{s}/v_{esc}=0.2913$. 
By simply multiplying the $K_{3}$ coefficient (see equation
\eqref{eq:scaling_law}) by $0.65$ (close to the ratio of the sound speed considered by
\citet{Reville:2014ud} and the one considered here), our results are
nicely reconciled with the predicted scaling-law (red dashed line and
red area). The power-law trend that
was derived from axisymmetric models with only three different
topologies \citep{Reville:2014ud} seems to apply, at least at first
order, to non-axisymmetric topologies involving numerous spherical harmonics modes.

\section{Conclusions}

We have presented a set of simulations to model the coronal structure
around distant stars based on observed magnetic maps. We applied 
this method to the case of HD 189733. Due to the lack
of constrains on the mass loss rate of the star, models of the stellar
wind of HD 189733 have free parameters, which we varied to explore their
impact on the coronal structure, and on the mass and angular momentum
loss rates of the star. In spite of the low
resolution we used, and the complexity of the magnetic field topology,
we find a good general agreement with the torque scaling law that was derived
in 2.5D geometry by \citet{Reville:2014ud} and our results compare
very well with previous models of the corona of HD 189733
\citep{Llama:2013il}. Nevertheless, the slight discrepancies compared to the predicted
torque scaling law warrant higher-resolution
runs. Furthermore, HD 189733 is a slow-rotator; we intend to confirm the torque scaling
law with three-dimensional simulations in the rapid-rotator regime as well in
a near future.

HD 189733 is known to harbor a close-in Jupiter-like planet, HD 189733b, which
orbits at 0.031~AU around its host \citep[$R_{p}=1.15\, R_{J}$ and
$M_{p} = 1.13\, M_{j}$, see][]{Boisse:2009aa}. It was recently
suggested that due to its proximity, HD 189733b could possess a
variable-size bow-shock along its phase, which could in principle be
detected in near-UV transit lights \citep{Llama:2013il}. Such
observations would then be extremely useful to better constrain
stellar wind models. The wind model we presented here is a first step
toward the global modelling of the star-planet system in the
spirit of \citet{Cohen:2014eb,Strugarek:2014ab}.

\begin{acknowledgements}
AS thank A. Vidotto for discussions about the modelling of the corona
of HD 189733. This work was supported by the ANR 2011 Blanc 
\href{http://ipag.osug.fr/Anr\_Toupies/}{Toupies}
and the ERC project 
\href{http://www.stars2.eu/}{STARS2} (207430). The
authors acknowledge CNRS INSU/PNST and CNES/Solar Orbiter
fundings. AS acknowledges support from the Canada's Natural Sciences and
Engineering Research Council and from the Canadian Institute of
Theoretical Astrophysics (National fellow). We acknowledge access to supercomputers
through GENCI (project 1623), Prace, and ComputeCanada infrastructures.
\end{acknowledgements}



\begin{thebibliography}{23}
\expandafter\ifx\csname natexlab\endcsname\relax\def\natexlab#1{#1}\fi

\bibitem[{Boisse {et~al.}(2009)Boisse, Moutou, Vidal-Madjar, Bouchy, Pont,
  H{\'e}brard, Bonfils, Croll, Delfosse, Desort, Forveille, Lagrange, Loeillet,
  Lovis, Matthews, Mayor, Pepe, Perrier, Queloz, Rowe, Santos, Segransan, \&
  Udry}]{Boisse:2009aa}
Boisse, I., Moutou, C., Vidal-Madjar, A., {et~al.} 2009, \aap, 495, 959

\bibitem[{Bouchy {et~al.}(2005)Bouchy, Udry, Mayor, Moutou, Pont, Iribarne,
  da~Silva, Ilovaisky, Queloz, Santos, Segransan, \& Zucker}]{Bouchy:2005aa}
Bouchy, F., Udry, S., Mayor, M., {et~al.} 2005, \aap, 444, L15

\bibitem[{Cohen \& Drake(2014)}]{Cohen:2014et}
Cohen, O. \& Drake, J.~J. 2014, \apj, 783, 55

\bibitem[{Cohen {et~al.}(2014)Cohen, Drake, Glocer, Garraffo, Poppenhaeger,
  Bell, Ridley, \& Gombosi}]{Cohen:2014eb}
Cohen, O., Drake, J.~J., Glocer, A., {et~al.} 2014, \apj, 790, 57

\bibitem[{Fares {et~al.}(2010)Fares, Donati, Moutou, Jardine, Grie{\ss}meier,
  Zarka, Shkolnik, Bohlender, Catala, \& Cameron}]{Fares:2010hq}
Fares, R., Donati, J.-F., Moutou, C., {et~al.} 2010, \mnras, 406, 409

\bibitem[{H{\'e}brard \& Lecavelier~des Etangs(2006)}]{Hebrard:2006aa}
H{\'e}brard, G. \& Lecavelier~des Etangs, A. 2006, \aap, 445, 341

\bibitem[{Llama {et~al.}(2013)Llama, Vidotto, Jardine, Wood, Fares, \&
  Gombosi}]{Llama:2013il}
Llama, J., Vidotto, A.~A., Jardine, M., {et~al.} 2013, \mnras, 2442

\bibitem[{Matt {et~al.}(2012)Matt, MacGregor, Pinsonneault, \&
  Greene}]{Matt:2012ib}
Matt, S.~P., MacGregor, K.~B., Pinsonneault, M.~H., \& Greene, T.~P. 2012,
  \apjl, 754, L26

\bibitem[{Mestel(1968)}]{Mestel:1968aa}
Mestel, L. 1968, \mnras, 138, 359

\bibitem[{Mignone {et~al.}(2007)Mignone, Bodo, Massaglia, Matsakos, Tesileanu,
  Zanni, \& Ferrari}]{Mignone:2007iw}
Mignone, A., Bodo, G., Massaglia, S., {et~al.} 2007, \apjs, 170, 228

\bibitem[{Mignone {et~al.}(2012)Mignone, Zanni, Tzeferacos, van Straalen,
  Colella, \& Bodo}]{Mignone:2012aa}
Mignone, A., Zanni, C., Tzeferacos, P., {et~al.} 2012, The Astrophysical
  Journal Supplement, 198, 7

\bibitem[{Oran {et~al.}(2013)Oran, van~der Holst, Landi, Jin, Sokolov, \&
  Gombosi}]{Oran:2013aa}
Oran, R., van~der Holst, B., Landi, E., {et~al.} 2013, \apj, 778, 176

\bibitem[{Parker(1958)}]{Parker:1958dn}
Parker, E.~N. 1958, \apj, 128, 664

\bibitem[{R{\'e}ville {et~al.}(2014)R{\'e}ville, Brun, Matt, \&
  Strugarek}]{Reville:2014ud}
R{\'e}ville, V., Brun, A.~S., Matt, S.~P., \& Strugarek, A. 2014, submitted to
  ApJ

\bibitem[{Schrijver \& DeRosa(2003)}]{Schrijver:2003vu}
Schrijver, C. \& DeRosa, M. 2003, \solphys, 212, 165

\bibitem[{Sokolov {et~al.}(2013)Sokolov, van~der Holst, Oran, Downs, Roussev,
  Jin, Manchester, Evans, \& Gombosi}]{Sokolov:2013ki}
Sokolov, I.~V., van~der Holst, B., Oran, R., {et~al.} 2013, \apj, 764, 23

\bibitem[{Strugarek {et~al.}(2012)Strugarek, Brun, \& Matt}]{Strugarek:2012th}
Strugarek, A., Brun, A.~S., \& Matt, S. 2012, in SF2A-2012: Proceedings of the
  Annual meeting of the French Society of Astronomy and Astrophysics. Eds.: S.
  Boissier, 419--423

\bibitem[{Strugarek {et~al.}(2014{\natexlab{a}})Strugarek, Brun, Matt, \&
  R{\'e}ville}]{Strugarek:2014fr}
Strugarek, A., Brun, A.~S., Matt, S.~P., \& R{\'e}ville, V. 2014{\natexlab{a}},
  Nature of Prominences and their role in Space Weather, 300, 330

\bibitem[{Strugarek {et~al.}(2014{\natexlab{b}})Strugarek, Brun, Matt, \&
  R{\'e}ville}]{Strugarek:2014tf}
Strugarek, A., Brun, A.~S., Matt, S.~P., \& R{\'e}ville, V. 2014{\natexlab{b}},
  th Cambridge Workshop on Cool Stars, Stellar Systems, and the Sun,
  Proccedings of Lowell Observatory, 1

\bibitem[{Strugarek {et~al.}(2014{\natexlab{c}})Strugarek, Brun, Matt, \&
  R{\'e}ville}]{Strugarek:2014ab}
Strugarek, A., Brun, A.~S., Matt, S.~P., \& R{\'e}ville, V. 2014{\natexlab{c}},
  \apj, 795, 86

\bibitem[{Weber \& Davis(1967)}]{Weber:1967kx}
Weber, E.~J. \& Davis, L.~J. 1967, \apjs, 148, 217

\bibitem[{Winn {et~al.}(2007)Winn, Holman, Henry, Roussanova, Enya, Yoshii,
  Shporer, Mazeh, Johnson, Narita, \& Suto}]{Winn:2007aa}
Winn, J.~N., Holman, M.~J., Henry, G.~W., {et~al.} 2007, The Astronomical
  Journal, 133, 1828

\bibitem[{Wood(2004)}]{Wood:2004uc}
Wood, B.~E. 2004, \lrsp, 1, 2

\end{thebibliography}

%
\end{document}